\documentclass[12pt,draftcls,peerreview,onecolumn]{IEEEtran}
\IEEEoverridecommandlockouts

\usepackage{amsmath}
\usepackage{amssymb}
\usepackage{graphicx}
\usepackage{psfrag}
\usepackage{url}
\usepackage{cite}

\graphicspath{{fig/}{num/}{adhocnet/}}
\newcommand{\E}{\mathrm{E}}

\DeclareMathOperator{\Tr}{Tr}
\DeclareMathOperator{\iid}{iid}
\DeclareMathOperator{\MIMO}{MIMO}
\DeclareMathOperator{\MAC}{MAC}
\DeclareMathOperator{\BC}{BC}
\DeclareMathOperator{\IC}{IC}

\DeclareMathOperator{\DPC}{DPC}

\DeclareMathOperator{\TX}{TX}
\DeclareMathOperator{\RX}{RX}
\DeclareMathOperator{\TXRX}{TX-RX}

\DeclareMathOperator{\tx}{tx}

\DeclareMathOperator{\NC}{NC}

\providecommand{\abs}[1]{\lvert{#1}\rvert}
\providecommand{\asq}[1]{\abs{#1}^2}

\begin{document}

\title{Capacity Gain from Two-Transmitter and Two-Receiver Cooperation}

\author{Chris~T.~K.~Ng,~\IEEEmembership{Student~Member,~IEEE,}
        Nihar~Jindal,~\IEEEmembership{Member,~IEEE,}\\
        Andrea~J.~Goldsmith,~\IEEEmembership{Fellow,~IEEE}
        and~Urbashi~Mitra,~\IEEEmembership{Fellow,~IEEE}
\thanks{This work was supported by the US Army under MURI award W911NF-05-1-0246, 
the ONR under award N00014-05-1-0168, DARPA's ITMANET program under grant 1105741-1-TFIND,
a grant from Intel, 
and the NSF ITR under grant CCF-0313392. 
The material in this paper was presented in part at the IEEE International Symposium on Information Theory, Chicago, IL, June/July 2004, and at the IEEE Information Theory Workshop, San Antonio, TX, October 2004.}%
\thanks{Chris~T.~K.~Ng and Andrea~J.~Goldsmith are with the Department of Electrical Engineering, Stanford University, Stanford, CA 94305 USA (e-mail: ngctk@wsl.stanford.edu; andrea@wsl.stanford.edu).}%
\thanks{Nihar~Jindal is with the Department of Electrical and Computer Engineering, University of Minnesota, Minneapolis, MN 55455 USA (e-mail: nihar@umn.edu).}%
\thanks{Urbashi~Mitra is with the Department of Electrical Engineering, University of Southern California,  Los Angeles, CA 90089 USA (e-mail: ubli@usc.edu).}%
}

\maketitle

\begin{abstract}

Capacity improvement from transmitter and receiver cooperation is investigated in a two-transmitter, two-receiver network with phase fading and full channel state information available at all terminals.
The transmitters cooperate by first exchanging messages over an orthogonal transmitter cooperation channel, then encoding jointly with dirty paper coding.
The receivers cooperate by using Wyner-Ziv compress-and-forward over an analogous orthogonal receiver cooperation channel.
To account for the cost of cooperation, the allocation of network power and bandwidth among the data and cooperation channels is studied.
It is shown that transmitter cooperation outperforms receiver cooperation and improves capacity over non-cooperative transmission under most operating conditions when the cooperation channel is strong.
However, a weak cooperation channel limits the transmitter cooperation rate; in this case receiver cooperation is more advantageous.
Transmitter-and-receiver cooperation offers sizable additional capacity gain over transmitter-only cooperation at low SNR, whereas at high SNR transmitter cooperation alone captures most of the cooperative capacity improvement.

\end{abstract}


\IEEEpeerreviewmaketitle

\section{Introduction}

\PARstart{I}{n} a wireless ad hoc network, each node can communicate with any other node over the wireless medium.
Hence groups of nodes may cooperate amongst one another to jointly encode or decode the transmission signals.
In this paper, we consider a scenario where there are two clustered transmitters and two clustered receivers, with each transmitter intending to send an independent message to a different receiver.
We assume the channels between the transmitter and receiver clusters are under quasi-static phase fading, and all terminals have perfect channel state information.
When the clustered nodes do not cooperate, the four-node network is an interference channel \cite{cover91:eoit}: its capacity remains an open problem in information theory.
We study this problem from a different perspective and ask the question:
How much does cooperation increase the set of achievable data rates?
However, we do not allow cooperation to occur for free.
We assume the nodes in a cluster cooperate by exchanging messages over an orthogonal cooperation channel which requires some fraction of the available network power and bandwidth.
To capture the cost of cooperation, we place a system-wide power constraint on the network, and examine different bandwidth allocation assumptions for the data and cooperation channels.

The notion of cooperative communication has been studied in several recent works.
In \cite{sendonaris03:coop1, sendonaris03:coop2} the authors showed that cooperation enlarges the achievable rate region in a channel with two cooperative transmitters and a single receiver.
Under a similar channel model, a non-orthogonal cooperative transmission scheme was presented in \cite{azarian03:coop_tx_schms}.
In \cite{hunter02:coop_coding}, the transmitters forward parity bits of the detected symbols, instead of the entire message, to achieve cooperation diversity.
A channel with two cooperative transmitters and two non-cooperative receivers was considered in terms of diversity for fading channels in \cite{laneman04:coop_diver}. It was shown that orthogonal cooperative protocols can achieve full spatial diversity order.
A similar channel configuration without fading was analyzed in \cite{host-madsen03:coop_rate} with dirty paper coding transmitter cooperation.
Achievable rate regions and capacity bounds for channels with transmitter and/or receiver cooperation were also presented in \cite{host-madsen06:coop_bounds, khojastepour04:coop_relay, host-madsen05:power_relay, host-madsen04:rx_coop, kramer05:coop_cap_relay, ng07:csi_pow_relay, host-madsen05:mplexg_gain, ng06:snr_mimo_rates}.

In this work, we examine the improvement in sum capacity from transmitter cooperation, receiver cooperation, and transmitter-and-receiver cooperation.
For transmitter cooperation, we consider dirty paper coding (DPC), which is capacity-achieving for multi-antenna Gaussian broadcast channels \cite{weingarten04:cap_mimo_bc}.
For receiver cooperation, we consider Wyner-Ziv compress-and-forward, which in relay channels is shown to be near-optimal when the cooperating node is close to the receiver \cite{kramer05:coop_cap_relay, ng07:csi_pow_relay}.
The dirty paper coding transmitter cooperation scheme was presented in \cite{jindal04:cap_coop, ng04:txcoop_dcp_relay}, with amplify-and-forward receiver cooperation in \cite{jindal04:cap_coop}.
Our work differs from previous research in this area in that i)~we consider receiver cooperation together with transmitter cooperation, and ii)~we characterize the cooperation cost in terms of power as well as bandwidth allocation in the network.

The remainder of the paper is organized as follows.
Section~\ref{sec:sys_mod} presents the system model.
In Section~\ref{sec:cap_gain_coop} we consider the gain in capacity from transmitter cooperation, receiver cooperation, and transmitter-and-receiver cooperation.
Numerical results of the cooperation rates, in comparison to corresponding multi-antenna channel capacity upper bounds, are presented in Section~\ref{sec:num_res}.
Section~\ref{sec:conclu} concludes the paper.

\section{System Model}
\label{sec:sys_mod}

\begin{figure}
  \centering
  \psfrag{G}[r][r]{$\sqrt{G}$}
  \psfrag{h1}[][]{$h_1$}
  \psfrag{h2}{$h_2$}
  \psfrag{h3}{$h_3$}
  \psfrag{h4}[tc][tc]{$h_4$}
  \psfrag{(x,y)}{$(x_i,y_i)$}
  \psfrag{(x1,y1)}{$(x_i',y_i')$}
  \psfrag{(x2,y2)}{$(x_i'',y_i'')$}
  \psfrag{Bt,Pt}{$B_t,P_t$}
  \psfrag{B,P1+P2}[r][r]{$B,P_1+P_2$}
  \psfrag{Br,Pr}{$B_r,P_r$}
  \includegraphics{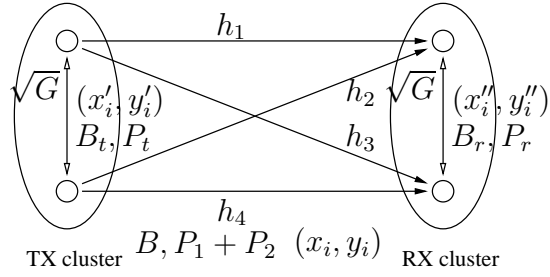}
  \caption{System model of a network with two clustered transmitters and two clustered receivers.}
  \label{fig:txrxcluster}
\end{figure}

Consider an ad hoc network with two clustered transmitters and two clustered receivers as shown in Fig.~\ref{fig:txrxcluster}.
We assume the nodes within a cluster are close together, but the distance between the transmitter and receiver clusters is large. 
The channel gains are denoted by $h_1,\dotsc,h_4\in\mathbb{C}$.
To gain intuition on the potential benefits of cooperation, we consider a simple model where the channels experience quasi-static phase fading \cite{kramer05:coop_cap_relay}: the channels have unit magnitude with independent identically distributed (iid) random phase uniform between 0 and $2\pi$.
Hence $h_i = e^{j\theta_i}$, $i=1,\dotsc,4$, where $\theta_i\sim \mathrm{U}[0,2\pi]$ and $\theta_i$ is fixed after its realization. 
We assume all nodes have perfect channel state information (CSI), and the transmitters are able to adapt to the realization of the channels. 

There are three orthogonal communication channels: the data channel between the transmitter and receiver clusters, the cooperation channel between the transmitters, and the cooperation channel between the receivers.
In the data channel, Transmitter~1 wishes to send to Receiver~1 at rate $R_1$, and likewise Transmitter~2 to Receiver~2 at rate $R_2$.
In this paper we investigate the capacity improvement in the sum rate $R_1+R_2$ from cooperation.
Let $\mathbf{x} \triangleq [x_1\;x_2]^T\in\mathbb{C}^2$ denote the transmit signals, and $\mathbf{y} \triangleq [y_1\;y_2]^T\in\mathbb{C}^2$ denote the corresponding received signals. In matrix form, the data channel can be written as
\begin{align}
  \mathbf{y} &= \mathbf{H} \mathbf{x} + \begin{bmatrix}n_1\\n_2\end{bmatrix}, &
  \mathbf{H}\triangleq \begin{bmatrix}h_1&h_2\\h_3&h_4\end{bmatrix},
\end{align}
where $n_1,n_2\sim\mathcal{CN}(0,1)$ are iid zero-mean circularly symmetric complex Gaussian (ZMCSCG) white noise with unit variance.
Let $B$ denote the bandwidth of the data channel, and $P_1 \triangleq \E[\asq{x_1}]$, $P_2 \triangleq \E[\asq{x_2}]$ denote the transmission power of Transmitter~1, Transmitter~2, respectively; the expectation is taken over repeated channel uses.

There is also a static, additive white Gaussian noise (AWGN) cooperation channel between the two transmitters with channel gain $\sqrt{G}$.
As we assume the cooperating nodes are close together, the case of interest is when $G$ is large. 
We assume the two transmitters can simultaneously transmit and receive on this full-duplex cooperation channel.
Let $x'_1,x'_2\in\mathbb{C}$ be the transmit signals, and $y'_1,y'_2\in\mathbb{C}$ the received signals, then the cooperation channel is described by
\begin{align}
  \label{eq:tx_coop_ch}
  y'_1 &= \sqrt{G}x'_2 + n_3, &
  y'_2 &= \sqrt{G}x'_1 + n_4,
\end{align}
where $n_3, n_4\sim\mathcal{CN}(0,1)$ are iid unit-variance ZMCSCG noise.
Let $B_t$ denote the transmitter cooperation channel bandwidth, and $P_t \triangleq \E\bigl[\asq{x'_1}+\asq{x'_2}\bigr]$ denote the cooperation transmission power.
Between the two receivers, there is an analogous full-duplex static AWGN cooperation channel, with channel gain equal to $\sqrt{G}$. 
Let $x''_1,x''_2\in\mathbb{C}$ be the signals sent on this channel, and $y''_1,y''_2\in\mathbb{C}$ be the received signals.
The receiver cooperation channel is then defined by
\begin{align}
  \label{eq:rx_coop_ch}
  y''_1 &= \sqrt{G}x''_2 + n_5, &
  y''_2 &= \sqrt{G}x''_1 + n_6,
\end{align}
where $n_5, n_6\sim\mathcal{CN}(0,1)$ are again iid unit-variance ZMCSCG noise.
Let $B_r$ denote the receiver cooperation channel bandwidth, and $P_r \triangleq \E\bigl[\asq{x''_1}+\asq{x''_2}\bigr]$ denote the receiver cooperation transmission power.

To capture the system-wide cost of transmitter and receiver cooperation, we consider a total network power constraint $P$ on the data and cooperation transmissions:
\begin{align}
  \label{eq:pow_constr}
  P_1 + P_2 + P_t + P_r \leq P.
\end{align}
Moreover, we consider two scenarios on the allocation of bandwidth between the data channel and the cooperation channels:
Under bandwidth assumption~1), we assume dedicated orthogonal channels for cooperation, and each channel has a bandwidth of 1~Hz (i.e., $B_t = B = B_r = 1$).
Under bandwidth assumption~2), however, there is a single 1~Hz channel to be divided into three different bands to implement the cooperative schemes.
We thus allocate $B_t,B$ and $B_r$ such that $B_t + B + B_r= 1$. 
Bandwidth assumption~1) is applicable when the short-range cooperation communications take place in separate bands which may be spatially reused across all cooperating nodes in the system, and hence the bandwidth cost for a particular cooperating pair can be neglected.
In contrast, bandwidth assumption~2) is applicable when spatial reuse is not considered.

\section{Capacity Gain from Cooperation}
\label{sec:cap_gain_coop}

\subsection{Transmitter Cooperation}
\label{sec:tx_coop}

We first consider transmitter cooperation in our network model assuming non-cooperating receivers (i.e., $P_r=0$, $B_r=0$).
In the transmitter cooperation scheme, the transmitters first fully exchange their intended messages over the orthogonal cooperation channel, after which the network becomes equivalent to a multi-antenna broadcast channel (BC) with a two-antenna transmitter:
\begin{align}
y_1 &= \mathbf{f}_1\mathbf{x} + n_1,&
y_2 &= \mathbf{f}_2\mathbf{x} + n_2,
\end{align}
where $\mathbf{f}_1, \mathbf{f}_2$ are the rows of $\mathbf{H}$:
\begin{align}
\mathbf{f}_1 &\triangleq \begin{bmatrix}h_1\;h_2\end{bmatrix},&
\mathbf{f}_2 &\triangleq \begin{bmatrix}h_3\;h_4\end{bmatrix}.
\end{align}
The transmitters then \emph{jointly} encode both messages using dirty paper coding (DPC), which is capacity-achieving for the multi-antenna Gaussian BC \cite{weingarten04:cap_mimo_bc}.
Causality is not violated since we can offset the transmitter cooperation and DPC communication by one block.
 
The sum capacity achieved by DPC in the multi-antenna BC is equal to the sum capacity of its dual multiple-access channel (MAC) \cite{vishwanath03:dual_mimo_bc, jindal04:dual_gaus_mabc}:
\begin{align}
R_{\DPC} & = B \log \lvert \mathbf{I} + \mathbf{f}_1^H(P_1/B)\mathbf{f}_1
+ \mathbf{f}_2^H(P_2/B)\mathbf{f}_2 \rvert\\
& = B \log\bigl(1 + 2(P_1+P_2)/B + 2\phi P_1P_2/B^2\bigr),
\end{align}
where $\log$ is base 2 and $\phi \triangleq 1-\cos(\theta_1-\theta_2-\theta_3+\theta_4)$. Note that $R_{\DPC}$ is symmetric and concave in $P_1,P_2$, thus it is maximized at $P_1^* = P_2^* = (P-P_t)/2$.
By symmetry each transmitter uses power $P_t/2$ to exchange messages in the cooperation channel, which supports the cooperation sum rate:
\begin{align}
 R_t & =2 B_t \log\bigl(1 + GP_t/(2B_t)\bigr).
\end{align}
To ensure each transmitter reliably decodes the other's message, we need $R_t\geq R_{\DPC}$; hence the transmitter cooperation sum rate is
\begin{align}
R_{\TX} &= \max_{B_t,B, 0\leq P_t \leq P} \; \min(R_t,R_{\DPC}).
\end{align}
Note that $R_t$ is increasing in $P_t$ while for $R_{\DPC}$ it is decreasing, the optimal $P_t^*$ is thus achieved at $R_t = R_{\DPC}$.
Under bandwidth assumption~1) with $B_t=B=1$, the optimal power allocation is
\begin{align}
P_t^* = \begin{cases}
\dfrac{2(\sqrt{D}-G-\phi P - 2)}{G^2-2\phi} & \text{if $G^2\neq 2\phi$}\\
\dfrac{P(\phi P + 4)}{2(G + \phi P + 2)} & \text{else},
\end{cases}
\end{align}
where $D \triangleq 4(G+1) + G^2(2P+1) + \phi GP(2+GP/2)$.
Under bandwidth assumption~2) with $B_t+B=1$, $P_t^*$ is found by equating $R_t$ and $R_{\DPC}$ for given $B_t,B$,
which is numerically computed as it involves solving equations with non-integer powers.
The optimal bandwidth allocation $B_t^*,B^*$ are found through numerical one-dimensional optimization \cite{forsythe77:com_math_comp}.

\subsection{Receiver Cooperation}
\label{sec:rx_coop}

Next we consider receiver cooperation in our network model without transmitter cooperation (i.e., $P_t=0$, $B_t=0$).
When the receivers cooperate, there is no advantage in using a decode-and-forward scheme since, due to channel symmetry, each receiver decodes just as well as its cooperating node does.
Instead, each receiver employs compress-and-forward to send a compressed representation of its own observation to the other receiver through the orthogonal cooperation channel.
The compression of the undecoded signal is realized using Wyner-Ziv source coding \cite{wyner76:rate_dist_side_info}, which exploits as side information the correlation between the observed signals of the receivers.
Compress-and-forward is shown to be near-optimal when the cooperating node is close to the receiver in relay channels \cite{kramer05:coop_cap_relay, ng07:csi_pow_relay}.

Suppose Receiver~2 sends its observation to Receiver~1 via compress-and-forward over the cooperation channel.
Then Receiver~1 is equivalent to a two-antenna receiver that observes the signals $[y_1\;y_2+z_2]^T$, where $z_2\sim\mathcal{CN}(0,\hat{N}_2)$ is the compression noise independent from $y_1,y_2$.
The variance of the compression noise is given in \cite{host-madsen06:coop_bounds, kramer05:coop_cap_relay}:
\begin{align}
\label{eq:comp_noise_2}
\hat{N}_2 &= \frac{\lvert\mathbf{H}(\Sigma_x/B)\mathbf{H}^H\rvert +
\mathbf{f}_2(\Sigma_x/B)\mathbf{f}_2^H + \mathbf{f}_1(\Sigma_x/B)\mathbf{f}_1^H + 1}
{(2^{R_{r2}/B}-1)\bigl(\mathbf{f}_1(\Sigma_x/B)\mathbf{f}_1^H+1\bigr)},
\end{align}
where $\Sigma_x \triangleq \E[\mathbf{xx}^H]$ is the covariance matrix of the transmit signals,
and $R_{r2}$ is the rate at which Receiver~2 compresses its observation with Wyner-Ziv source coding.
Likewise, Receiver~1 follows similar compress-and-forward procedures to send its observation to Receiver~2 at rate $R_{r1}$ with compression noise $z_1$ that has variance $\hat{N}_1$ given by:
\begin{align}
\label{eq:comp_noise_1}
\hat{N}_1 &= \frac{\lvert\mathbf{H}(\Sigma_x/B)\mathbf{H}^H\rvert +
\mathbf{f}_2(\Sigma_x/B)\mathbf{f}_2^H + \mathbf{f}_1(\Sigma_x/B)\mathbf{f}_1^H + 1}
{(2^{R_{r1}/B}-1)\bigl(\mathbf{f}_2(\Sigma_x/B)\mathbf{f}_2^H+1\bigr)}.
\end{align}
In the absence of transmitter cooperation (i.e., $\Sigma_x$ is diagonal), note that the compression noise variance is symmetric in Receiver~1 and Receiver~2.
Suppose Receiver~1, Receiver~2 use power $Pr_1,Pr_2$, respectively, to perform compress-and-forward, then the Wyner-Ziv source coding rate is given by the capacity of the receiver cooperation channel:
\begin{align}
\label{eq:rx_co_Rri}
R_{ri} = B_r \log\bigl(1+GP_{ri}/B_r)\bigr),\quad i=1,2.
\end{align}

When each receiver has a noisy version of the other's signal, the network is equivalent to a multi-antenna interference channel where each receiver has two antennas:
\begin{align}
\mathbf{\tilde{y}}_1 &= \mathbf{\tilde{h}}_1 x_1 + \mathbf{\tilde{h}}_2 x_2
+ \begin{bmatrix}n_1\; \tilde{n}_2\end{bmatrix}^T\\
\mathbf{\tilde{y}}_2 &= \mathbf{\tilde{h}}_3 x_1 + \mathbf{\tilde{h}}_4 x_2
+ \begin{bmatrix}\tilde{n}_1\; n_2 \end{bmatrix}^T,
\end{align}
with
\begin{align}
\mathbf{\tilde{h}}_1 &\triangleq \begin{bmatrix}h_1 \; \sqrt{\eta_2}h_3 \end{bmatrix}^T, &
\mathbf{\tilde{h}}_2 &\triangleq \begin{bmatrix}h_2 \; \sqrt{\eta_2}h_4\end{bmatrix}^T,\\
\mathbf{\tilde{h}}_3 &\triangleq \begin{bmatrix}\sqrt{\eta_1} h_1\; h_3\end{bmatrix}^T, &
\mathbf{\tilde{h}}_4 &\triangleq \begin{bmatrix}\sqrt{\eta_1} h_2\; h_4\end{bmatrix}^T,
\end{align}
$\mathbf{\tilde{y}}_i$ for $i=1,2$ is the aggregate signal from reception and cooperation at Receiver~$i$, $\tilde{n}_i \sim \iid \mathcal{CN}(0,1)$, and $\eta_i \triangleq 1/(1+\hat{N_i})$ is the degradation in antenna gain due to the compression noise.
We assume equal power allocation $P_{r1}=P_{r2}=P_r/2$ between the receivers, which results in the symmetric compress noise variance:
\begin{align}
\hat{N}_1 = \hat{N}_2 \triangleq \hat{N}
=\frac{2\phi P_1P_2/B^2 + 2(P_1+P_2)/B + 1}
{\bigl([1+GP_r/(2B_r)]^{B_r/B}-1\bigr)\bigl((P_1+P_2)/B + 1\bigr)}.
\end{align}
Equal power allocation achieves the saddle point that satisfies the strong interference condition \cite{sato81:gaus_strong_interf}, under which each receiver decodes the messages from both transmitters and symmetric allocation of the receiver cooperation power is optimal. The sum capacity of the interference channel is
\begin{align}
\begin{split}
R_{\IC} &= \min \bigl\{ B \log \lvert \mathbf{I} + \mathbf{\tilde{h}}_1 (P_1/B) \mathbf{\tilde{h}}_1^H 
+ \mathbf{\tilde{h}}_2 (P_2/B) \mathbf{\tilde{h}}_2^H \rvert, \\
 & \qquad \quad B \log \lvert \mathbf{I} + \mathbf{\tilde{h}}_3 (P_1/B) \mathbf{\tilde{h}}_3^H 
+ \mathbf{\tilde{h}}_4 (P_2/B) \mathbf{\tilde{h}}_4^H \rvert \bigr\}
\end{split}\\
&= B \log\bigl(1+(1+\eta)(P_1+P_2)/B + 2\eta\phi P_1P_2/B^2 \bigr),
\end{align}
where $\eta_1=\eta_2\triangleq\eta$.
The interference channel sum capacity is symmetric in $P_1,P_2$, but not concave.
As $R_{\IC}$ does not have a structure that lends readily to analytical maximization,
the receiver cooperation sum rate is found through numerical exhaustive search over the power and bandwidth allocation variables:
\begin{align}
R_{\RX} &= \max_{B,B_r} \; R_{\IC}\quad\text{subject to:} \; P_r \leq P - P_1 - P_2.
\end{align}

\subsection{Transmitter and Receiver Cooperation}
\label{sec:txrx_coop}

The cooperation schemes described in the previous sections can be combined by having the transmitters exchange their messages and then perform DPC, while the receivers cooperate using compress-and-forward.
Let $\mathcal{C}_{\tx}(G,P_t,B_t)$ denote the rate region supported by the transmitter cooperation channel, with $(R_1,R_2) \in \mathcal{C}_{\tx}(G,P_t,B_t)$ iff
\begin{align}
(2^{R_1/B_t}-1)B_t/G + (2^{R_2/B_t}-1)B_t/G \leq P_t,
\end{align}
which follows from the capacity of AWGN channels.
Suppose Receiver~1 uses power $P_{r1}$ to compress-and-forward to Receiver~2 with compression noise $z_1\in\mathcal{CN}(0,\hat{N}_1)$, and in the opposite direction Receiver~2 uses power $P_{r2}$ with compression noise $z_2\in\mathcal{CN}(0,\hat{N}_2)$,
where $\hat{N}_2, \hat{N}_1$ are as given in (\ref{eq:comp_noise_2}), (\ref{eq:comp_noise_1}).

When both transmitters know the intended transmit messages, and each receiver has a noisy version of the other's signal, the network is equivalent to a multi-antenna BC with a two-antenna transmitter and two two-antenna receivers:
\begin{align}
\mathbf{\tilde{y}}_1 &= \mathbf{\tilde{H}}_1 \mathbf{x} + \begin{bmatrix}n_1 \; \tilde{n}_2\end{bmatrix}^T\\
\mathbf{\tilde{y}}_2 &= \mathbf{\tilde{H}}_2 \mathbf{x} + \begin{bmatrix}\tilde{n}_1\; n_2 \end{bmatrix}^T,
\end{align}
where
\begin{align}
\label{eq:H_tilde_1}
\mathbf{\tilde{H}}_1 &\triangleq \begin{bmatrix}h_1 & h_2 \\ \sqrt{\eta_2}h_3 & \sqrt{\eta_2}h_4\end{bmatrix},&
\eta_2 &\triangleq 1/(1+\hat{N}_2)\\
\label{eq:H_tilde_2}
\mathbf{\tilde{H}}_2 &\triangleq \begin{bmatrix}\sqrt{\eta_1}h_1 & \sqrt{\eta_1}h_2 \\ h_3 & h_4 \end{bmatrix},&
\eta_1 &\triangleq 1/(1+\hat{N}_1).
\end{align}
Suppose the transmit signals intended for Receiver~1, Receiver~2 have covariance matrices $\Sigma_{x1},\Sigma_{x2}$, respectively.
Note that $\Sigma_x = \Sigma_{x1} + \Sigma_{x2}$ since DPC yields statistically independent transmit signals.
Let DPC encode order~$(1)$ denote encoding for Receiver~1 first, then for Receiver~2; the corresponding DPC rates are
\begin{align}
R_{1,\DPC}^{(1)} &= B \log \frac{\lvert\mathbf{I}+\mathbf{\tilde{H}}_1(\Sigma_x/B)\mathbf{\tilde{H}}_1^H\rvert}
{\lvert\mathbf{I}+\mathbf{\tilde{H}}_1(\Sigma_{x2}/B)\mathbf{\tilde{H}}_1^H\rvert}\\
R_{2,\DPC}^{(1)} &= B \log \lvert\mathbf{I}+\mathbf{\tilde{H}}_2(\Sigma_{x2}/B)\mathbf{\tilde{H}}_2^H\rvert.
\end{align}
Under encode order~$(2)$, when DPC encoding is performed for Receiver~2 first followed by Receiver~1, the rates are
\begin{align}
R_{1,\DPC}^{(2)} &= B \log \lvert\mathbf{I}+\mathbf{\tilde{H}}_1(\Sigma_{x1}/B)\mathbf{\tilde{H}}_1^H\rvert\\
R_{2,\DPC}^{(2)} &= B \log \frac{\lvert\mathbf{I}+\mathbf{\tilde{H}}_2(\Sigma_x/B)\mathbf{\tilde{H}}_2^H\rvert}
{\lvert\mathbf{I}+\mathbf{\tilde{H}}_2(\Sigma_{x1}/B)\mathbf{\tilde{H}}_2^H\rvert}.
\end{align}
The transmitter-and-receiver cooperation sum rate is given by the following optimization problem:
\begin{align}
  R_{\TXRX} &= \max_{B_t,B,B_r} \; R_1 + R_2\\
  \text{subject to:} \;
(R_1,R_2)&\in\mathcal{C}_{\tx}(G,P_t,B_t)\\
(R_1,R_2)&\in \bigl\{(R_{1,\DPC}^{(i)},R_{2,\DPC}^{(i)}),\; i=1,2 \bigr\}\\
\Tr(\Sigma_{x1} + \Sigma_{x2}) &\leq P-P_t-P_{r1}-P_{r2}.
\end{align}
In general, the power and bandwidth allocated for transmitter cooperation and receiver cooperation need to be optimized jointly. 
However, since the search space is large, we consider a suboptimal allocation scheme.
We assume a fixed compression noise target ($\hat{N}_1, \hat{N}_2$) that is supported by receiver cooperation, with which the equivalent multi-antenna BC matrices are $\mathbf{\tilde{H}}_1, \mathbf{\tilde{H}}_2$ as given in (\ref{eq:H_tilde_1}), (\ref{eq:H_tilde_2}).
Then we find the optimal transmitter input distributions $\Sigma_{x1}^*,\Sigma_{x2}^*$ that maximize the DPC BC sum rate for $\mathbf{\tilde{H}}_1, \mathbf{\tilde{H}}_2$ using the sum power iterative waterfilling algorithm \cite{jindal05:sum_pow_iter_wf}.
In the end we verify through bisection search that the total power required to achieve the DPC rates and to support ($\hat{N}_1, \hat{N}_2$) is feasible under the network power constraint.
The transmitter-and-receiver cooperation sum rate is then found through numerical exhaustive search over ($\hat{N}_1, \hat{N}_2$) and the bandwidth allocation variables.
In the numerical results the suboptimal allocation scheme is able to achieve rates that approach the MIMO capacity upper bound as $G$ increases.

\section{Numerical Results}
\label{sec:num_res}

In this section we present the capacity gain achieved by the cooperation schemes at different SNRs under different bandwidth assumptions.
The cooperation rates are compared against the baseline of non-cooperative sum capacity.
With neither transmitter nor receiver cooperation, the network is a Gaussian interference channel
under strong interference, and the non-cooperative sum capacity is
\begin{align}
  C_{\NC} &= \min\bigl\{ \log(1+\asq{h_1}P_1+\asq{h_2}P_2),
             \,\log(1+\asq{h_3}P_1+\asq{h_4}P_2) \bigr\}\\
&= \log(1+P).
\end{align}
The cooperation rates are also compared to the capacity of corresponding multi-antenna channels as if the cooperating nodes were colocated and connected via a wire.
With such colocated transmitters, the channel becomes a multi-antenna BC with a two-antenna transmitter. The BC sum capacity is given by the sum capacity of its dual MAC:
\begin{align}
\label{eq:C_BC_sum}
C_{\BC} &= \max_{P_1+P_2\leq P}\;
\log \lvert \mathbf{I} + \mathbf{f}_1^H P_1\mathbf{f}_1
+ \mathbf{f}_2^H P_2\mathbf{f}_2 \rvert\\
& = \log\bigl(1 + 2P + \phi P^2/2\bigr),
\end{align}
where the last equality follows from the sum capacity being symmetric and concave in $P_1,P_2$: thus it is maximized at $P_1^*=P_2^*=P/2$.
With similarly colocated receivers, the channel becomes a multi-antenna MAC with a two-antenna receiver, the sum capacity of which is given by:
\begin{align}
\label{eq:C_MAC_sum}
C_{\MAC} & = \max_{P_1+P_2\leq P}\; \log \lvert \mathbf{I} + \mathbf{h}_1 P_1\mathbf{h}_1^H
+ \mathbf{h}_2 P_2\mathbf{h}_2^H \rvert\\
\label{eq:C_MAC_sum_P_sym}
& = \log\bigl(1 + 2P + \phi P^2/2\bigr),
\end{align}
where $\mathbf{h}_1,\mathbf{h}_2$ are the columns of $\mathbf{H}$: 
\begin{align}
\mathbf{h}_1 &\triangleq [h_1 \; h_3]^T, &
\mathbf{h}_2 &\triangleq [h_2 \; h_4]^T,
\end{align}
and (\ref{eq:C_MAC_sum_P_sym}) follows again from the symmetry and concavity of $P_1,P_2$.
Note that the MAC in (\ref{eq:C_MAC_sum}) is not the dual channel of the BC in (\ref{eq:C_BC_sum}); nonetheless they evaluate to the same sum capacity: $C_{\BC}=C_{\MAC}$.
Last, with colocated transmitters \emph{and} colocated receivers, the channel becomes a MIMO channel where the transmitter and the receiver each have two antennas. The capacity of the MIMO channel is
\begin{align}
C_{\MIMO} &= \max_{\Tr(\Sigma_x)\leq P} \log \lvert \mathbf{I} + \mathbf{H}\Sigma_x\mathbf{H}^H\rvert,
\end{align}
where the optimal input covariance matrix $\Sigma_x^*$ is found by waterfilling over the eigenvalues of the channel \cite{telatar99:cap_mimo_gaus}.

\begin{figure}[p]
  \centering
  \includegraphics*[width=8cm]{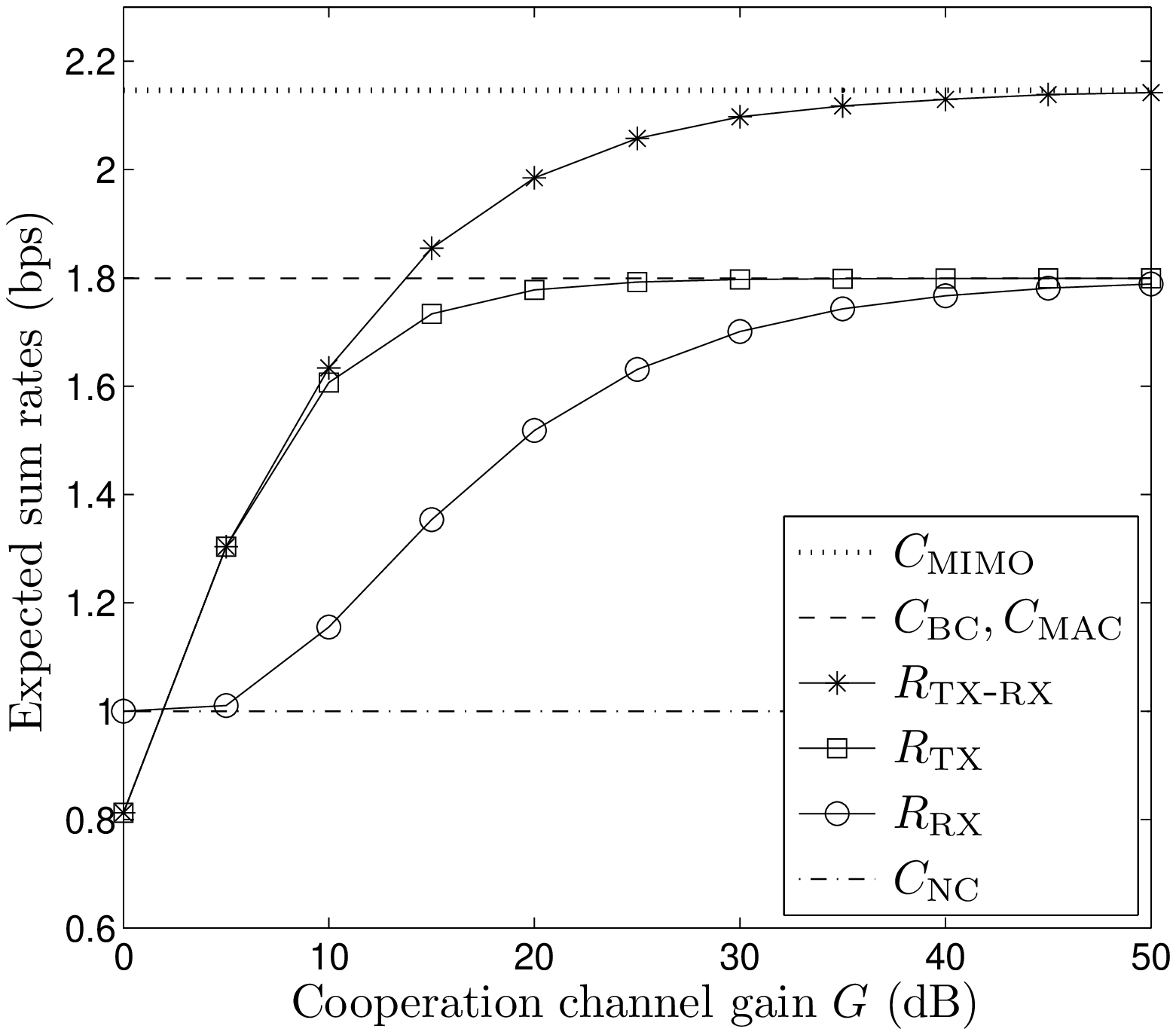}
  \caption{Cooperation sum rates under bandwidth assumption~1) at $P=0$~dB.}
  \label{fig:coop_rates_0dB_BW1}
\end{figure}

\begin{figure}
  \centering
  \includegraphics*[width=8cm]{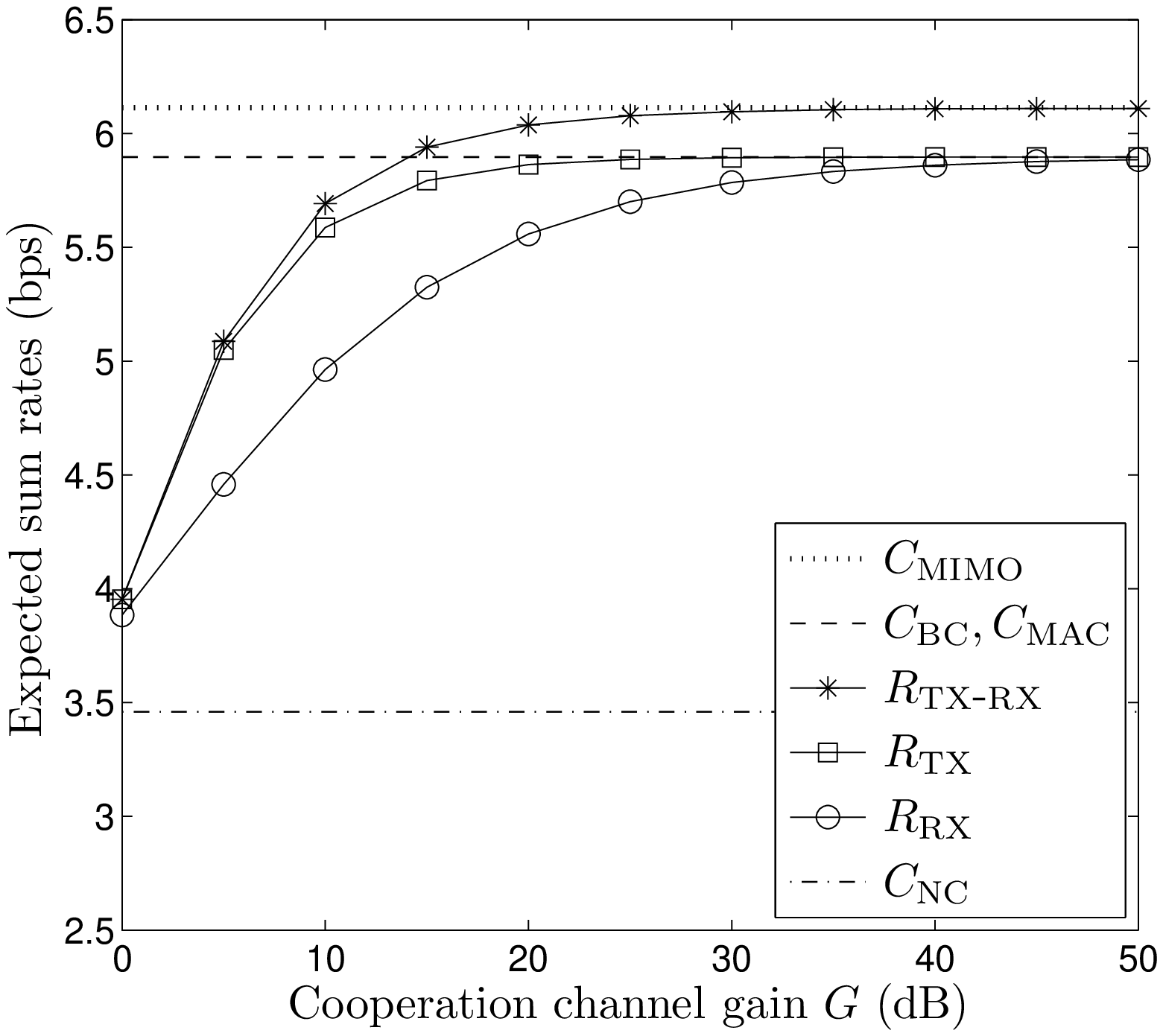}
  \caption{Cooperation sum rates under bandwidth assumption~1) at $P=10$~dB.}
  \label{fig:coop_rates_10dB_BW1}
\end{figure}

\begin{figure}
  \centering
  \includegraphics*[width=8cm]{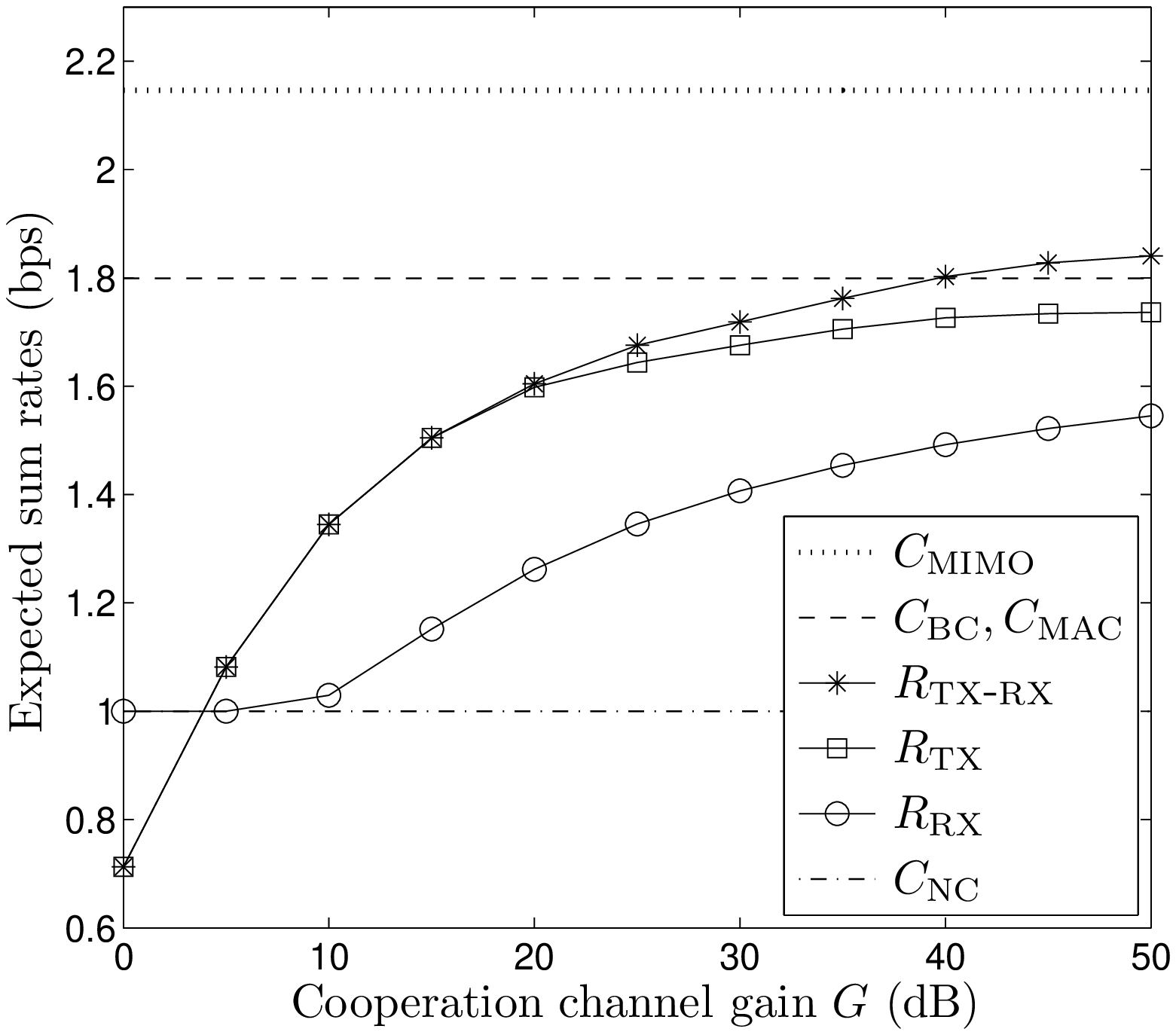}
  \caption{Cooperation sum rates under bandwidth assumption~2) at $P=0$~dB.}
  \label{fig:coop_rates_0dB_BW2}
\end{figure}

\begin{figure}
  \centering
  \includegraphics*[width=8cm]{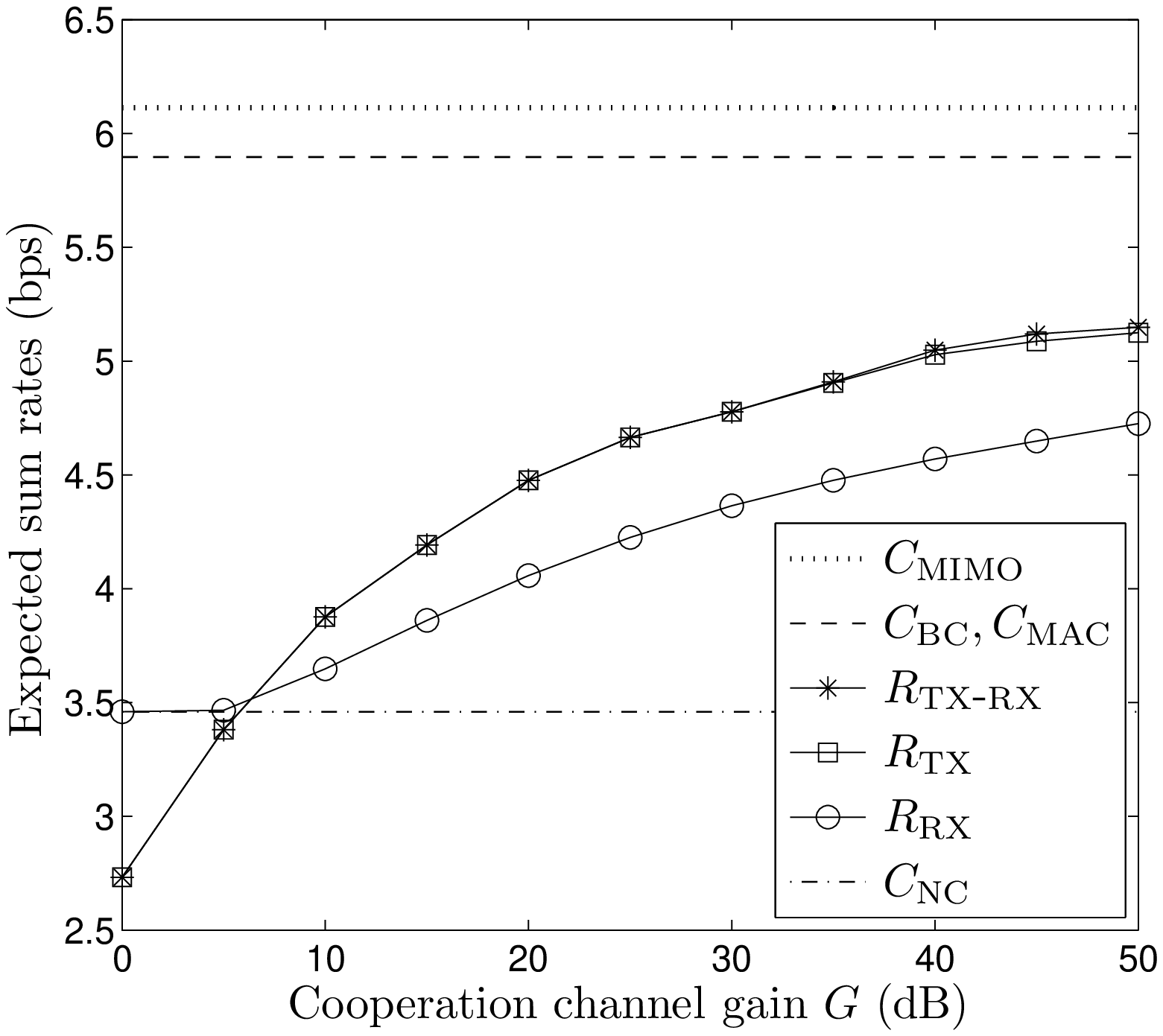}
  \caption{Cooperation sum rates under bandwidth assumption~2) at $P=10$~dB.}
  \label{fig:coop_rates_10dB_BW2}
\end{figure}

Fig.~\ref{fig:coop_rates_0dB_BW1} and Fig.~\ref{fig:coop_rates_10dB_BW1} show the cooperation rates at SNRs of $P=0$~dB and 10~dB, respectively, under bandwidth assumption~1) where the cooperation channels occupy separate dedicated bands.
Fig.~\ref{fig:coop_rates_0dB_BW2} and Fig.~\ref{fig:coop_rates_10dB_BW2} show the cooperation rates at 0~dB and 10~dB under bandwidth assumption~2) where the network bandwidth is allocated among the data and cooperation channels.
The expected rates are computed via Monte Carlo simulation over random channel realizations.
Under bandwidth assumption~2), as the network bandwidth needs to be divided among the data and cooperation channels, it takes a higher cooperation channel gain $G$ to achieve cooperation rates comparable to those under bandwidth assumption~1); nevertheless, under both bandwidth assumptions the relative performance of transmitter and receiver cooperation follows similar trends.
When $G$ is small, the transmitter cooperation rate is impaired by the provision that each transmitter decodes the message of the other, which becomes a performance burden under a weak cooperation channel.
Receiver cooperation, on the other hand, always performs better than or as well as non-cooperative transmission, since the compress-and-forward rates adapt to the channel conditions.
When $G$ is higher than approximately 5~dB, however, the receiver cooperation rate begins to trail behind that of transmitter cooperation.

When $G$ is large, both $R_{\TX}$ and $R_{\RX}$ approach the multi-antenna channel capacity $C_{\BC}$ and $C_{\MAC}$.
For the range of $G$ when the cooperation rates are far below the BC or MAC capacity, transmitter-and-receiver cooperation offers minimal capacity improvement over transmitter-only cooperation.
As $G$ increases further, $R_{\TX}$ and $R_{\RX}$ are bounded by $C_{\BC}$ and $C_{\MAC}$, but $R_{\TXRX}$ continues to improve and approaches the MIMO channel capacity $C_{\MIMO}$.
As SNR increases, however, the additional capacity improvement from transmitter-and-receiver cooperation over transmitter-only cooperation becomes insignificant, as $C_{\BC}$ tends to $C_{\MIMO}$ in the limit of high SNR \cite{caire03:ach_guas_mimo_bc}.

\section{Conclusion}
\label{sec:conclu}

We have studied the capacity improvement in the sum rate from DPC transmitter cooperation, Wyner-Ziv compress-and-forward receiver cooperation, as well as transmitter-and-receiver cooperation when the cooperating nodes form a cluster in a two-transmitter, two-receiver network with phase fading and full channel state information available at all terminals.
To account for the cost of cooperation, we imposed a system-wide transmission power constraint, and considered the allocation of power and bandwidth among the data and cooperation channels.
It was shown that transmitter cooperation outperforms receiver cooperation and improves capacity over non-cooperative transmission under most operating conditions when the cooperation channel is strong.
However, when the cooperation channel is very weak, it becomes a bottleneck and transmitter cooperation underperforms non-cooperative transmission; in this case receiver cooperation, which always performs at least as well as non-cooperation, is more advantageous.
Transmitter-and-receiver cooperation offers sizable additional capacity gain over transmitter-only cooperation at low SNR, whereas at high SNR transmitter cooperation alone captures most of the cooperative capacity improvement.

We considered a simple model where the channels between the transmitter and receiver clusters are under phase fading to gain intuition on the potential benefits of cooperation.
When the channels are under Rayleigh fading, for example, power and bandwidth allocation become less tractable since we cannot exploit the symmetry in the channels; however, the DPC and compress-and-forward cooperation schemes are still applicable and we expect comparable cooperative capacity gains can be realized.
We assumed perfect CSI are available at all terminals; the system model is applicable in slow fading scenarios when the channels can be tracked accurately.
The CSI assumption is critical: without CSI we cannot perform DPC or Wyner-Ziv compression effectively and we expect the benefits of cooperation to be considerably diminished.


\bibliographystyle{ieeebib/IEEEtran.bst}
\bibliography{ieeebib/IEEEabrv,bib/wrlscomm}

\end{document}